\documentclass[iop,revtex4]{emulateapj} 

\usepackage{graphicx,subfigure}
\usepackage{float}
\usepackage{booktabs}
\usepackage{commath}
\usepackage{amsmath}
\usepackage{color}

\newcommand{\comm}[1]{}

\shorttitle{Artificial, incoherent calibration speckles}
\shortauthors{N. Jovanovic et al.}

\begin{document}
\title{Artificial incoherent speckles enable precision astrometry \\and photometry in high-contrast imaging}
\author{N. Jovanovic\altaffilmark{1,2}, O. Guyon\altaffilmark{1,3,4}, F. Martinache\altaffilmark{5}, P. Pathak\altaffilmark{1}, J. Hagelberg\altaffilmark{6}, and T. Kudo\altaffilmark{1}}
\altaffiltext{1}{National Astronomical Observatory of Japan, Subaru Telescope, 650 North A'Ohoku Place, Hilo, HI, 96720, U.S.A.}
\altaffiltext{2}{Department of Physics and Astronomy, Macquarie University, NSW 2109, Australia}
\altaffiltext{3}{Steward Observatory, University of Arizona, Tucson, AZ, 85721, U.S.A.}
\altaffiltext{4}{College of Optical Sciences, University of Arizona, Tucson, AZ 85721, U.S.A.}
\altaffiltext{5}{Observatoire de la Cote d'Azur, Boulevard de l'Observatoire, Nice, 06304, France}
\altaffiltext{6}{Institute for Astronomy, University of Hawaii, 2680 Woodlawn Drive, Honolulu, HI 96822, U.S.A.}
\email{jovanovic.nem@gmail.com}

\begin{abstract}
State-of-the-art coronagraphs employed on extreme adaptive optics enabled instruments, are constantly improving the contrast detection limit for companions at ever closer separations to the host star. In order to constrain their properties and ultimately compositions, it is important to precisely determine orbital parameters and contrasts with respect to the stars they orbit. This can be difficult in the post coronagraphic image plane, as by definition the central star has been occulted by the coronagraph. We demonstrate the flexibility of utilizing the deformable mirror in the adaptive optics system in SCExAO to generate a field of speckles for the purposes of calibration. Speckles can be placed up to $22.5~\lambda/D$ from the star, with any position angle, brightness and abundance required. Most importantly, we show that a fast modulation of the added speckle phase, between $0$ and $\pi$, during a long science integration renders these speckles effectively incoherent with the underlying halo. We quantitatively show for the first time that this incoherence in turn, increases the robustness and stability of the adaptive speckles which will improve the precision of astrometric and photometric calibration procedures. This technique will be valuable for high-contrast imaging observations with imagers and integral field spectrographs alike. 
\end{abstract}


\section{Introduction}
The field of high-contrast imaging has received a major boost to its arsenal in recent times with the completion of commissioning of GPI~\citep{macintosh14} and SPHERE~\citep{bez08}, which join P$1640$~\citep{hinkley11}. These systems optimized for high-contrast imaging close to the host star share similar underlying architectures: they exploit an extreme adaptive optics system (ExAO) which stabilizes the point-spread function (PSF) before suppressing the starlight with coronagraphs to reveal a faint companion. 

Key to constraining the properties of the companion, including its atmospheric composition, is the ability to accurately determine its position and distance from the host star and its relative brightness as a function of wavelength. This was recognized early on and a solution which utilizes a diffractive grid placed in the pupil of the telescope, that generates fixed speckles in the focal plane was implemented on P$1640$~\citep{Opp06}. Owing to its simplicity and robustness, GPI has more recently also exploited this concept~\citep{wang14,macintosh14}. Each speckle generated by the grid is a miniature replica of the central PSF with a pre-determined contrast (usually $>100:1$), allowing the faint companion and speckles to be captured within the limited dynamic range of the camera. Most importantly, satellite speckles allow for the exact position of the PSF to be triangulated behind the coronagraph. Finally, as the speckles are replicas of the host star, they can be used to collect a spectrum of the host. This implementation has been successfully used on-sky to characterize binary systems~\citep{zim10} and improve the orbital properties of the well known planet $\beta$ Pic b~\citep{macintosh14}.

However, this method has several shortcomings. Firstly, the diffractive grid consists of opaque lines, which for example could be marked on a plate~\citep{macintosh14}, which diffract the light by modulating the amplitude in the pupil. This results in a measurable reduction in throughput. Secondly, since the grid is fixed, so are the speckle positions and contrasts. For these reasons the more recent Subaru Coronagraphic Extreme Adaptive Optics (SCExAO) system, utilizes phase modulated diffractive grids generated by its deformable mirror (DM) instead~\citep{jovanovic15}. 

Regardless of which technology is used, the grids generate speckles that are coherent with the residual speckle halo around the PSF - resulting in interference which leads to deformation of the reference speckles ultimately limiting calibration precision. However, by alternating the phase of the speckles between $0$ and $\pi$ several times during an exposure, time-averaged ``incoherent" speckles which are immune to the underlying speckle halo can be realized. This offers the highest possible precision for calibrating the photometry/astrometry of the companion. A DM is ideal in this case as it is relatively simple to modulate the phase in this way, but could also be implemented with a diffractive pupil mask by simply scanning the mask laterally with respect to the beam. In this work we outline for the first time, the process of generating incoherent calibration speckles and demonstrate the successful implementation in simulations and on-sky. 

\section{Principle}
To explain how the grid generates speckles in the focal plane, we can study an analogous system - the diffraction grating. A diffraction grating can be understood by imagining each groove in the grating as a source, and applying Huygens principle to determine in which direction the waves interfere destructively/constructively. The paths of constructive interference constitute the diffraction orders. Since, the diffracted beams have angular diversity (i.e. they are not parallel), then by focusing them with an optic an image with multiple spots will be generated. In this way, a periodic diffracting component (whether it be a grid mask or a DM) placed in the pupil plane can be used to generate speckles in the focal plane. 

Much like a diffraction-grating, one can modify the grid to control the placement of the speckles. For example, the separation between the PSF and the speckles is controlled by the period of the grid with respect to the pupil size. More periods across the pupil project the grid further from the PSF. For the $45$ actuators across the SCExAO DM, the furthest speckles can be projected is $22.5~\lambda/D$ from the PSF~\citep{jovanovic15} ($900$~mas in H-band) (for P1640 it is $32~\lambda/D$ as it has $64$ actuators across the pupil~\citep{Opp13}). In addition, the thickness of the lines if using a transmission mask or the amplitude of the sine wave on the DM will determine the brightness of the speckles (i.e. how much light is to be diffracted). The rotational orientation of the sine wave in the pupil can be used to control the orientation of the speckles around the PSF. Unlike, a diffraction grating however, the grids used for astrometry do not have a so-called ``blaze angle". They are simply uniform lines - power is coupled evenly amongst the orders symmetrically about the PSF. Finally, the phase of the speckles can be varied by translating the grid across the pupil. This final property is key to generating incoherent speckles. 

To understand how incoherent speckles can be formed, we must first examine the equation that governs interference between two beams given by
\begin{multline}\label{eq:interference}
\norm{ A_{h}e^{i\phi}+A_{s}e^{i\theta}}^{2}=A_{h}^{2}+A_{s}^{2}+2A_{h}A_{s}cos(\phi+\theta)
\end{multline}
where $A_{h}$ and $\phi$ represent the amplitude and phase of the underlying speckle in the halo surrounding the PSF and $A_{s}$ and $\theta$ represent the amplitude and phase of the artificial speckle applied. The first two terms on the right side of Equation~\ref{eq:interference}, correspond to the incoherent sum of the individual speckle intensities. The third term, depends on the amplitude and phase of the speckles. If we now assume an image is taken at time $t=0$ with an artificial speckle applied with a phase of $0$, and then a subsequent image is taken shortly after at time $t=0+\delta$ with the phase of the artificial speckle modulated by $\pi$ (while the amplitude remains constant) we can describe the intensity at a given point in the focal plane as
\begin{align}
t=0:                I_{1} &= A_{h1}^{2}+A_{s1}^{2}+2A_{h1}A_{s1}cos(\phi_{1})\label{eq:interference2}\\
t=0+\delta:    I_{2} &= A_{h2}^{2}+A_{s2}^{2}+2A_{h2}A_{s2}cos(\phi_{2}+\pi)\label{eq:interference3}\\
		                 &= A_{h2}^{2}+A_{s2}^{2}-2A_{h2}A_{s2}cos(\phi_{2}).\nonumber
\end{align}
If the time interval is sufficiently short such that the amplitude and phase of the halo speckle can be assumed constant (i.e. $A_{h1}=A_{h2}=A_{h}$ and $\phi_{1}=\phi_{2}=\phi$), then if the images were averaged together the final intensity for the speckle is given by
\begin{align}\label{eq:interference4}
(I_{1}+I_{2})/2 		&= (2A_{h}^{2}+2 A_{s}^{2}+2A_{h}A_{s}cos(\phi)\\&-2A_{h}A_{s}cos(\phi))/2\nonumber\\
				&= A_{h}^{2}+A_{s}^{2}.\nonumber
\end{align}
Hence, by averaging two images, closely spaced in time such that the properties of the speckles in the halo are fixed, while modulating the phase of the artificial speckle by $\pi$, it is possible to get a resultant speckle that only depends on the amplitude of the two individual speckles interacting, which is essentially ``incoherent". As a final step, $A_{h}^{2}$ can be removed by subtracting off a reference PSF image leaving only $A_{s}^{2}$, the artificial speckle intensity to calculate photometry/astrometry. We will show this powerful property results in more stable speckles which will be beneficial for high precision astrometric/photometric calibration. 

\section{Simulation}\label{sec:sim}
\begin{figure}
\centering 
\includegraphics[width=0.9\linewidth]{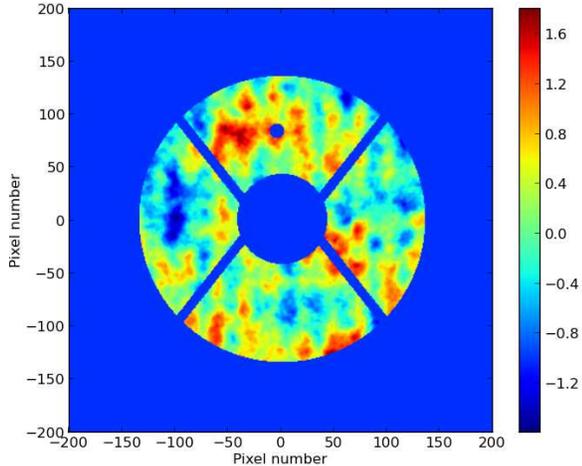}
\caption{\footnotesize Subaru pupil geometry with a Kolmogorov phase screen and a sine wave. The spot in the upper sector of the pupil corresponds to the location of a dead actuator on the SCExAO DM. The colorbar is in units of radians.}
\label{fig:pupil}
\end{figure}

\begin{figure}
\centering 
\includegraphics[width=0.99\linewidth]{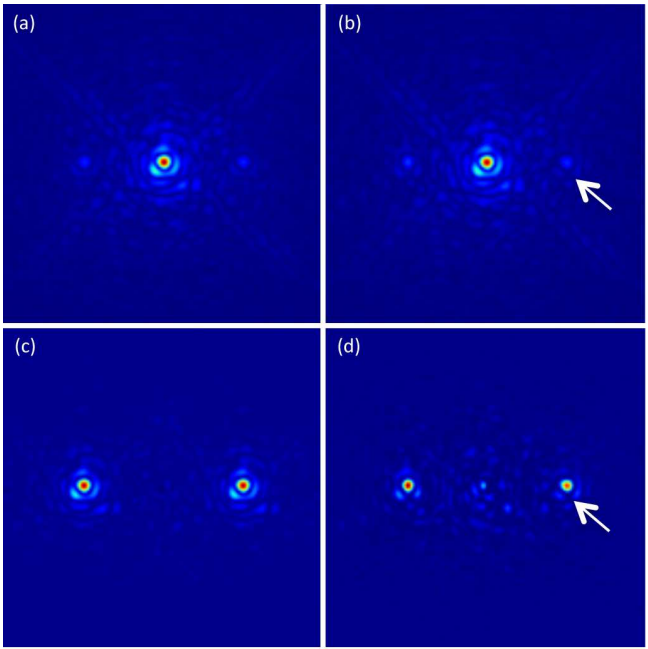}
\caption{\footnotesize (Top) PSF with $2$ artificial speckles at $10~\lambda/D$ from the PSF. (a) Incoherent speckles. (b) Coherent speckles. (Bottom) PSF subtracted image (c) with incoherent speckles (d) with coherent speckles.}
\label{fig:simulations}
\end{figure}

To simulate the effect of artificial coherent and incoherent speckles, the Subaru Telescope pupil geometry was used and realistic wavefronts (phase fronts) were added. For clarity we define incoherent speckles as those generated by the average of two images with an artificial speckle applied of constant amplitude and a phase of $0$ in one image and $\pi$ in the other, while coherent speckles consist of only a single speckle phase ($0$). Figure~\ref{fig:pupil} shows a simulated Kolmogorov phase screen where the low spatial frequencies were attenuated to mimic a low-order AO correction (typical of most observatories these days, RMS of $0.6$ radians), with the addition of a sine wave ($10$ cycles across the pupil). The corresponding focal plane image with coherent speckles can be seen in the top right panel of Fig.~\ref{fig:simulations}. A second image was generated with the same sine wave simply phase shifted by $\pi$. The two images were averaged together and the resultant image with incoherent speckles is shown in the top left panel of Fig.~\ref{fig:simulations}. A third image was generated without the sine wave applied (a PSF reference), and subtracted from the images with the speckles. Note, in this body of work only a straightforward subtraction of the reference PSF from the images with speckles was used - no advanced algorithms such as LOCI~\citep{Lafreniere07} or KLIP~\citep{soummer12} were implemented, the impact of which will be discussed in Section~\ref{sec:osexp}. The resultant images with speckles in isolation are shown in the bottom of Fig.~\ref{fig:simulations}.

The top panels in Fig.~\ref{fig:simulations} nicely demonstrate the typical speckle halo post-AO correction which includes: diffraction from the spiders, Airy rings, residual turbulent speckles as well as the two deliberate artificial speckles (which are at a similar brightness). It can be seen from the PSF subtracted image of the incoherent speckles that the artificial speckles are truly a replica of the PSF (see the similarly scaled PSF in the top left panel). In addition, the incoherent speckles are also identical to one another, unlike the coherent ones. Indeed, both the shape of the speckles and the first Airy rings of the two coherent speckles look different. The white arrow in the right hand set of panels indicates one obvious modulation of the Airy ring which is brought about by interference with the speckle background and hence not seen in the incoherent set of speckles. Finally, it can be seen that there is more noise in the PSF subtracted image of the coherent speckles which is a result of speckle noise. It is clear that the incoherent speckles are robust against modulation and deformation due to the speckle background and hence will offer superior calibration. 

\section{On-sky validation}\label{sec:osexp}
The SCExAO instrument is a testbed for state-of-the-art technologies and is optimized for high-contrast imaging at $<3~\lambda/D$. It combines ExAO, infrared coronagraphy and high-angular resolution visible interferometry. A full discussion of the instrument was presented by~\cite{jovanovic15}, here we only highlight the key features pertinent to this work. SCExAO utilizes a $2000$ element DM ($45$ illuminated actuators across the beam) which is conjugated to the telescope pupil. Downstream, the light is split and sent to a $320\times256$ pixel InGaAs detector which is used to monitor the PSF in the NIR at high speed, while deeper exposures are taken with the H2RG in the HiCIAO camera~\citep{suzuki10}. For laboratory testing a calibration source is used which is based on a super continuum source. 

On-sky testing was conducted on Beta Leo (spectral type$=A3$, H-mag$=1.92$) on the SCExAO engineering night of the $1^{st}$ of April $2015$. Data was collected with the upstream facility AO systems (AO$188$) loop closed offering Strehls between $20-40\%$ (H-band). The pyramid wavefront sensor was not operated on that night. There were patchy clouds overhead, the seeing was $\sim0.5"$ (H-band) and the airmass was $\sim1.07$ when the data was acquired. Speckles were generated with $100$~nm RMS amplitudes for each sine wave on the DM and projected at $10~\lambda/D$ from the PSF in both the vertical and horizontal directions. The separation was chosen to position the speckles in the residual AO halo deliberately in order to demonstrate the concept more clearly. A $50$~nm filter centered at $1600$~nm, was placed in front of the InGaAs detector to restrict the bandwidth and minimize elongation due to chromaticity of speckles, for simplicity. The integration time was set to prevent saturation ($5$~ms). Initially, a cube of $1000$ images was taken with the speckles statically applied to the DM. A second cube of $1000$ images was taken with the phase of both sine waves modulated between $0$ and $\pi$ every $10$~ms. As will be shown the $10$~ms switching time was sufficiently fast to demonstrate the concept discussed here. Finally, a cube without any speckles was collected for PSF subtraction purposes. 

\begin{figure}
\centering 
\includegraphics[width=0.99\linewidth]{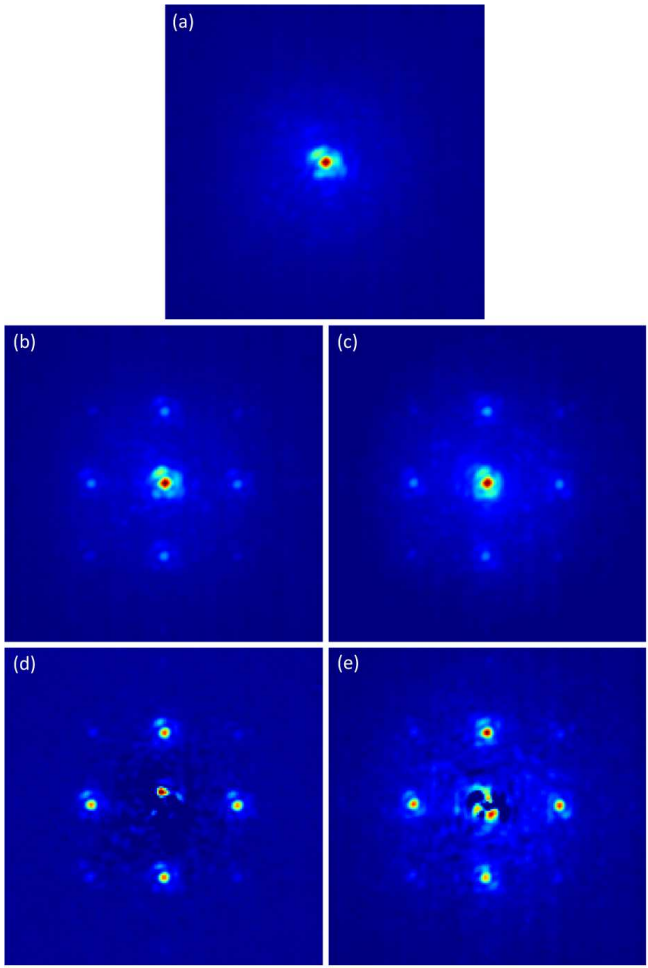}
\caption{\footnotesize (a) Image of PSF. PSF with $2$ sets of artificial speckles at $10~\lambda/D$ ($400$~mas) from the PSF, (b) incoherent speckles, (c) coherent speckles. PSF subtracted image (d) with incoherent speckles (e) with coherent speckles. A square-root stretch was applied and the minimum and maximum of each image adjusted for maximum contrast. Data taken on Beta Leo on the $1^{st}$ of April, $2015$.}
\label{fig:onsky}
\end{figure}

A mean dark frame, calculated from a cube of $1000$ darks, was subtracted from each frame in the science cubes. Hot pixels were removed. As the phase of the sine waves was modulated at a lower frequency and not synchronized with the acquisition time of the camera, images were co-added to simulate the affect of an incoherent set of speckles. The data was binned into $50$/$100$-frame bins (downsampling to $20$/$10$ data points respectively, by shifting and averaging). These bin sizes correspond to temporal windows of $0.5-1$~s which are of the order of the shortest exposures typically taken with H2RG detectors in the NIR. The reference PSF was subtracted from both binned data sets. A single averaged-image of the PSF, PSF with speckles applied, as well as the incoherent speckles and coherent speckles post PSF subtraction can be seen in Figure~\ref{fig:onsky}. 

It can be seen by comparing panels (d) and (e) of Figure~\ref{fig:onsky} that there is a dramatic difference in the speckles between the incoherent and coherent cases. The incoherent speckles are both more similar in shape and in brightness from qualitative inspection. To quantify this the photometry of each speckle was extracted. This was done by calculating the encircled flux after precise background subtraction. The standard deviation of the photometric signal for each speckle along the cubes was calculated, and then they were averaged to offer a single value for the stability of the photometry of the speckles in each cube. The results are summarized in Table~\ref{tab:results}. 

\begin{deluxetable}{lccc}[h!]
\tabletypesize{\footnotesize}
\centering
\tablecaption{Comparison of photometric and astrometric precision \\with incoherent and coherent speckles respectively. \label{tab:results}}
\tablehead{
\colhead{}				& 	\colhead{$50$-frame bin} 	&	\colhead{$100$-frame bin} }	\\
\startdata      
\bf{Photometry}			&	(Average fractional RMS)	&							\\
Coherent	  			& 	$0.30$				&	$0.25$					\\
Incoherent			&	$0.15$				&	$0.11$					\\
\bf{Astrometry}			&   	(pixels)				&							\\   
Coherent				&      $0.45$				& 	$0.30$		  			\\
Incoherent			&	$0.17$				&	$0.13$					\\
\enddata
\tablecomments{The photometry displayed in the table corresponds to the average RMS value for all four speckles for each data set. The astrometry refers to the average RMS value of the cross-correlations between each speckle.}
\end{deluxetable}

The astrometric precision was determined by cross-correlating the speckles with one another. In an ideal case where there is no speckle modulation due to interference, one would expect the cross-correlation to be constant along a cube, i.e. the distance between two speckles remains constant with time and hence the RMS of the cross-correlation would approach zero. Firstly, the standard deviation of each cross-correlation vector (between two speckles) was calculated along the cube and then the average of the standard deviation across all vectors was computed and is summarized in Table~\ref{tab:results}. 

It can be seen from Table~\ref{tab:results} that there is an improvement of $2$--$3\times$ in the stability of the photometry/astrometry of the grid when incoherent speckles were implemented. This is an important improvement which will directly result in the ability to constrain the properties of companions to a greater precision when used in conjunction with advanced PSF subtraction techniques, enabling more accurate modeling of system dynamics and atmospheric compositions (through photometric variability) and their structures to be undertaken. For completeness the average speckle contrast with respect to the PSF was determined to be $6.4\pm1.1\%$ in this case.     

\section{Considerations}\label{sec:consider}
There are several terms which will contribute to the error budget for determining the precision of the astrometry/photometry for a companion. The technique proposed here aims at reducing the error in the pixel-to-sky registration and photometric calibration terms (i.e. a grid thats well referenced to the sky is created forming a yardstick to calibrate the companions properties). However, the error that comes from extracting the companions PSF from the speckle background is not addressed. This is the job of advanced PSF subtraction techniques (e.g. LOCI/KLIP). The improvement provided by the technique proposed here will depend on which term dominates the error budget. The gird stability measurement conducted on-sky clearly validated that the first terms described above were indeed addressed by the modulated grid of spots. Assuming however, that the PSF subtraction is handled well, then the technique proposed here will be beneficial and the aspects listed below should be considered.  

The first consideration is the speed used to do the phase switch. Ideally, the phase would be switched in $<1$~ms (or as fast as possible) which would ensure that the amplitude/phase of the halo speckle remains constant across each pair of phase modulated images and hence that the resultant speckle is purely incoherent. As long as each pair of speckles is incoherent it is possible to bin the images or take a long exposure which will preserve the incoherent nature of the speckles with improved signal-to-noise. Also, if discrete phase modulation is not possible, by for example needing to translate a grid mask across the pupil, then the total switch time must be kept as short as possible. 

The second consideration is the reproducibility of the brightness of the artificial speckles. This is typically limited by the lack of open-loop calibration of the deflection as a function of applied voltage of the DM. If the DM is not calibrated then the artificial speckle pair may not have equal amplitudes which will lead to an increase in the standard deviavtion of the speckle photometry. More importantly though, terms $3$ and $4$ in Equation~\ref{eq:interference4} will not cancel as $A_{s1}\neq A_{s2}$ and hence the speckles will no longer be incoherent, reducing calibration precision. 

In the case that both considerations can not be met, then this leads to the third consideration which is the relative brightness and location of the residual speckles with respect to the artificial speckles. For typical observations the brightness of the artificial speckles is chosen such that they can be captured along with the scientific target within the dynamic range of the detector. ExAO systems are ambitiously pushing for a raw contrast of $10^{5}$ at small angular separations ($<500$~mas). At such contrasts the speckles must be carefully positioned to reduce the affect of the non-canceling cross-terms in Equation~\ref{eq:interference4}. ExAO systems readily offer $90\%$ Strehl in the NIR now, form a dark hole around the PSF and improve the residual speckle contrast to the $10^{3}$ level at small angular separation. In the case that such a system used its DM to generate the modulated grid of spots described here, the ideal position for the artificial spots would be inside the dark hole where there is the greatest suppression of the background. If however, a diffractive pupil mask were used as is the case for GPI, then the lowest speckle background can be found at larger separations from the PSF than the control region of the DM permits, which is a result of the decreasing power of atmospheric turbulence with increasing spatial frequency. Regardless of technology, the placement of the speckles should not interfere with the scientifically interesting region around the PSF. This third consideration will maximize performance given imperfectly incoherent speckles. As a note, the separation of the artificial speckle from the host need not match that of the companion.  

Finally, the precision of both the astrometry/photometry can be enhanced by utilizing a greater number of speckles. The precision will improve with the $\sqrt{n}$ ($n$ is the number of independent sine waves used). This trend will only be true if the deflection/voltage relationship is well calibrated for the DM. In addition, by utilizing a larger number of speckles the distortions in the optical train downstream of the grid plane can also be calibrated. More speckles however means less flux in the companion making it fainter which needs to be taken into consideration. 

\section{Summary}\label{sec:summary}
In this body of work we demonstrate the application of artificial incoherent speckles to the focal plane of a high-contrast imaging instrument and the resultant improvement in the stability of the grid that is formed by this method. This technique can be used with both grid masks (by translating them with respect to the pupil) or deformable mirrors. The technique is independent of the imager used and will work equally well for integral field spectrographs. Further investigation into how the modulated grid of spots interacts with advanced PSF subtraction techniques and the resultant performance should be undertaken. This will help resolve questions about the optimum observing mode (constantly modulating grid vs on/off grid cycling), amongst others. Such a method will no doubt lead to more precise constraints placed on substellar companions which will help to refine dynamic stability models as well as offer details about exoplanet atmospheric compositions/structures. 

\acknowledgments
Hagelberg is supported by the Swiss National Science Foundation (SNSF). The authors acknowledge support from the JSPS (Grant-in-Aid for Research \#$23340051$ \& \#$26220704$). 

\textit{Facilities: Subaru.}

\end{document}